\documentclass{ws-procs9x6}

\begin{document}

\title{Interaction of the $\Theta^+$ with the nuclear medium}

\author{M.~J. VICENTE VACAS, D. CABRERA, Q.~B. LI, V.~K. MAGAS and E. OSET}

\address{ Departamento de F\'{\i}sica Te\'orica and IFIC,\\ 
Centro Mixto  Universidad de Valencia - CSIC, \\
Institutos de Investigaci\'on de Paterna, \\ 
 Apdo. correos 22085, 46071, Valencia, Spain}

\maketitle

\abstracts{We study the selfenergy of the $\Theta^+$ pentaquark in nuclei
associated with two types of interaction: the $KN$ decay of the $\Theta^+$
and  two meson baryon decay channels of the $\Theta^+$. Whereas the potential
related to the first source is quite weak, the second kind of interaction
produces a large and attractive potential that could lead to the existence of
$\Theta^+$ nuclear bound states.}

\section{Introduction}

 The discovery of the $\Theta^+$ resonance\cite{Nakano:2003qx} opens the
possibility of forming exotic $\Theta^+$ hypernuclei, which, like in the case of
negative strangeness, could provide complementary information on the structure
and properties of the pentaquark to that obtained from elementary reactions.

 In Ref. \refcite{Miller:2004rj}, a schematic model for quark-pair interaction
with nucleons was used to describe the $\Theta^+$, which suggested that
$\Theta^+$ hypernuclei, stable against strong decay, might exist. Later, in
Ref. \refcite{Kim:2004fk}, the $\Theta^+$ selfenergy in the nuclei is
calculated, based on the $\Theta^+ K N$ interaction. The results show
a selfenergy  too weak to bind the $\Theta^+$ in nuclei.

In this talk, we present some selected results from Ref. 
\refcite{Cabrera:2004yg}. There, we redo the calculations of Ref. 
\refcite{Kim:2004fk} modifying some of the assumptions.  The results are
similar to those of Ref. \refcite{Kim:2004fk} and a quite small potential is
obtained from this source.  Additionally, we consider other new selfenergy
pieces related to the coupling of the $\Theta^+$ resonance to a  baryon  and 
two mesons under certain assumptions. We find that the in-medium
renormalization of the  two meson cloud of the $\Theta^+$ could lead to a
sizable attraction, enough to produce bound and narrow $\Theta^+$ states in
nuclei.

The coupling of the $\Theta^+$ to two mesons and a nucleon is studied using  a
$SU(3)$ symmetric Lagrangian\cite{madrid}, constructed to account for the
coupling of the antidecuplet to a baryon and two pseudoscalar mesons.  With
this  Lagrangian an attractive selfenergy is obtained for all the members of
the antidecuplet coming from the two meson cloud.

\section{The $\Theta^+$ selfenergy in nuclear matter}

\subsection{Selfenergy from the $KN$ decay channel} 

\begin{figure}
\begin{center} 
    \includegraphics[height=5.0cm]{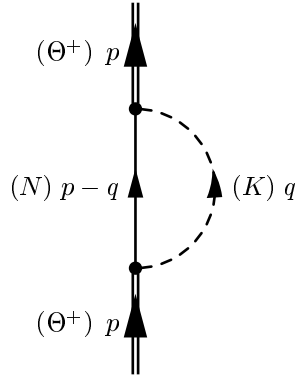}
\end{center} 
\caption{$\Theta^+$  selfenergy diagram related to the $K N$ decay channel.}
\label{2bodyself}
\end{figure}

The $\Theta^+$  selfenergy diagram in vacuum is shown in Fig. \ref{2bodyself}.
For the $L=0$ case, the free $\Theta^+$ selfenergy from this diagram is given 
by
\begin{equation}
-i\Sigma_{KN}=2\int\frac{d^4q}{(2\pi)^4} g_{K^+n}^2 
\frac{M}{E_N}
\frac{1}{p^0-q^0-E_N+i\epsilon}
\frac{1}{q^2-m_K^2+i\epsilon}\,,
\end{equation}
where $M$ is the nucleon mass, $E_N(k)=\sqrt{M^2+\vec{k}\,^2}$. The results 
for $L=1$ are obtained by the substitution 
$g_{K^+n}^2\rightarrow \bar{g}_{K^+n}^2 \vec{q}\,^2 \, .$
  
 Next, we evaluate the $\Theta^+$  selfenergy in  infinite nuclear
matter with density $\rho$. The nucleon propagator is changed in the
following way,
\begin{equation}
\frac{1}{p^0-q^0-E_N+i\epsilon}\rightarrow 
\frac{1-n(\vec{p}-\vec{q})}{p^0-q^0-E_N+i\epsilon}\,+\,
 \frac{n(\vec{p}-\vec{q})}{p^0-q^0-E_N-i\epsilon}\,,
\end{equation}
where $n(\vec{k})$ is the occupation number. The vacuum kaon propagator 
is replaced by the in-medium one,
\begin{equation}
\frac{1}{q^2-m_K^2+i\epsilon}\rightarrow\frac{1}{q^2-m_K^2-\Pi_K(q,\rho)}\,,
\end{equation}
where $\Pi_K(q^0,|\vec{q}|,\rho)$ is the kaon selfenergy. The $s-$wave part 
of this selfenergy is well approximated by\cite{Kaiser:1996js,Oset:2000eg} 
$
\Pi_{K}^{(s)}(\rho)=0.13\, m_K^2 \rho/\rho_0\ \,,
$
where $\rho_0$ is the normal nuclear density. The $p-$wave part is taken from 
the model of Ref. \refcite{Cabrera:2002hc},
which accounts for $\Lambda h$, $\Sigma h$ and $\Sigma^*(1385) h$ excitations.
After some approximations, the $q^0$ integration leads to
\begin{eqnarray}
\label{SigmaAll}
& &\Sigma_{KN} (p^0,\vec{p};\rho) = 
\nonumber \\
&=& \frac{M_{\Theta} \Gamma}{M q_{on}} 
\int \frac{d^3 q}{(2\pi)^2}  \frac{M}{E_N} {F}_L (q)
\frac{1}{2\widetilde{\omega}(q)} \,
\frac{1}{p^0-\widetilde{\omega}(q)-E_N-V_N+i\epsilon}
\nonumber \\
&-& \frac{M_{\Theta} \Gamma}{M q_{on}} 
\int \frac{d^3 q}{(2\pi)^2} \frac{M}{E_N} {F}_L (q)
\frac{n(\vec{p}-\vec{q})}
{(q^0)^2-\vec{q}\,^2-m_K^2-\Pi_K}
\end{eqnarray}
with $q^0=p^0-E_N(\vec{p}-\vec{q})-V_N$, $V_N=-\frac{k_F^2}{2M}$,  ${
F}_0=1$, ${ F}_1=\frac{\vec{q}\,^2}{q_{on}^2}$ and $q_{on}$ the on shell
kaon momentum.   In Eq. \ref{SigmaAll} we
have  taken into account the nucleon binding by using the Thomas-Fermi 
potential, $V_N=-k_F(r)^2/2M$. For the calculations, we have taken an 
average value of the momentum of the $\Theta^+$ in eventual bound states of 
$p=200$ MeV, similar to that of bound nucleons in nuclei.

\begin{figure}
\begin{center} 
   \includegraphics[width=0.85\textwidth]{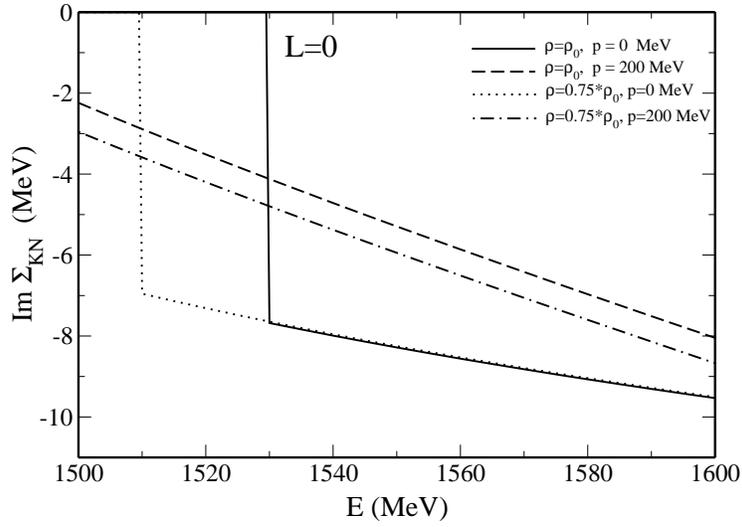}
\end{center} 
\caption{Imaginary part of the $\Theta^+$ selfenergy associated to the $KN$ 
decay channel for $L=0$.}
\label{ImL0}
\end{figure}

\begin{figure}
\begin{center} 
    \includegraphics[width=0.85\textwidth]{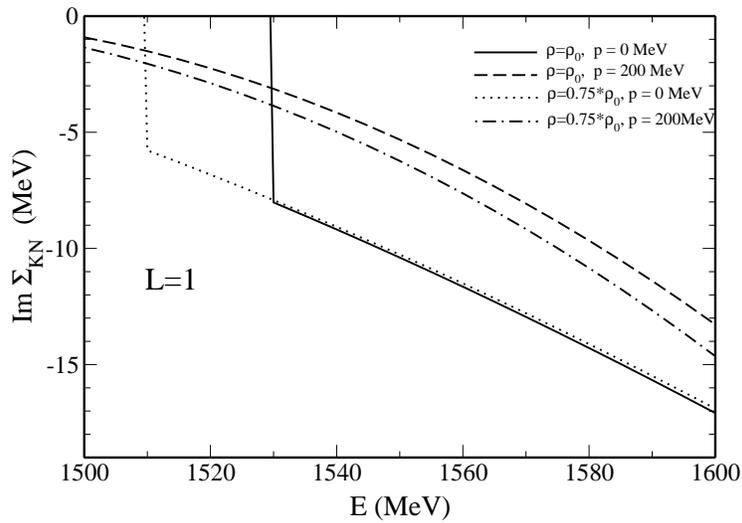}
\end{center} 
\caption{Imaginary part of the $\Theta^+$ selfenergy associated to 
the $KN$ decay channel for  $L=1$.}
\label{ImL1}
\end{figure}

 We show the results in Figs. \ref{ImL0}, \ref{ImL1}, where we assume  that
$\Gamma=15$ MeV.  As it is obvious from Eq. \ref{SigmaAll}, the in medium
selfenergy  is proportional to the vacuum width, thus the results should be 
scaled when the width is better determined. 
 In any case, even if $\Gamma=15$ MeV in vacuum, inside the nucleus the width 
is small, basically because of the Pauli blocking. For $20$ MeV of $\Theta^+$
binding, the width would go down from $15$ MeV to less than $6$ MeV
at $\rho=\rho_0$. This width
could be  smaller than the separation between different bound levels.
As for the real part of the $\Theta^+$ selfenergy, we find,
in qualitative agreement with Ref. \refcite{Kim:2004fk}, that the $\Theta^+$
potential in the medium is small, of the order of 1 MeV  or less and not enough
to bind $\Theta^+$ in nuclei.

\subsection{The $\Theta^+$ selfenergy tied to the two-meson cloud}
\begin{figure}
\begin{center} 
\includegraphics[width=0.7\textwidth]{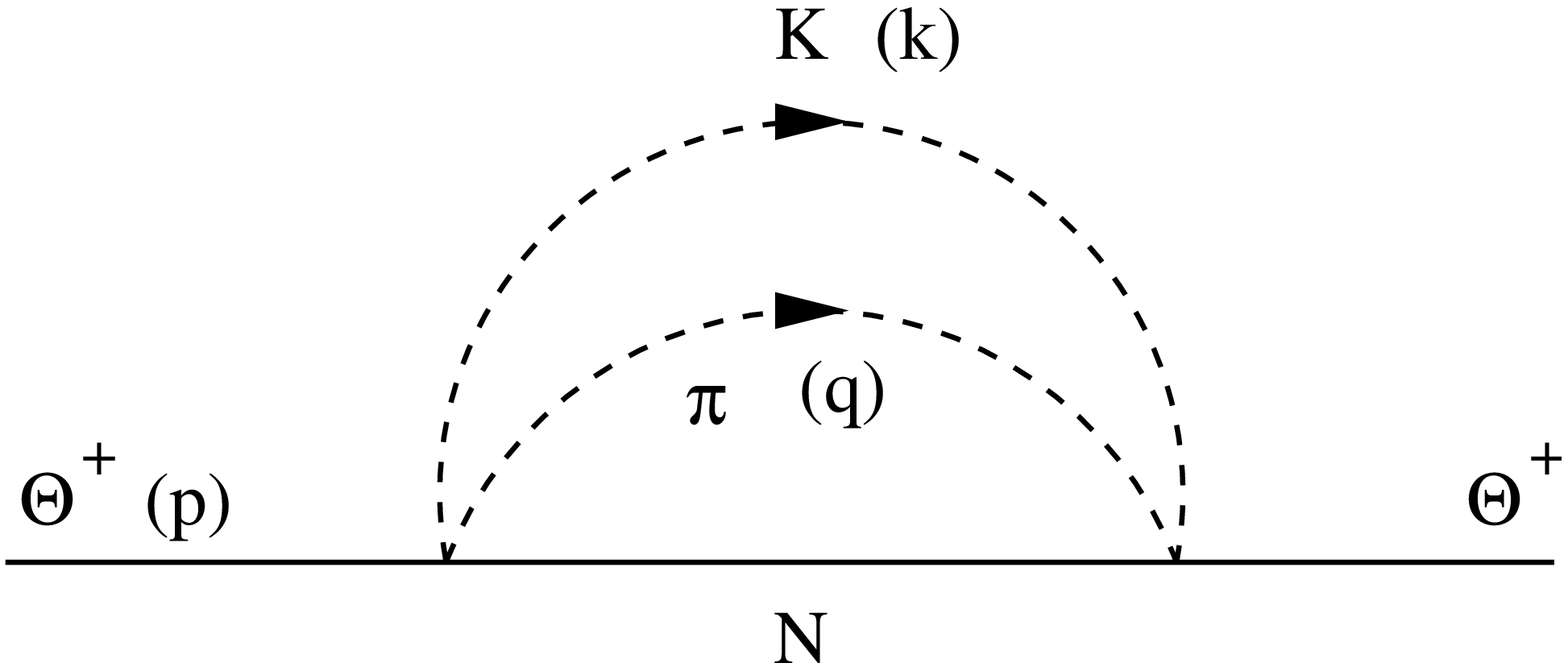}
\end{center} 
\caption{\label{Threebody} Two-meson $\Theta^+$ selfenergy diagram.}
\end{figure}

In this section
we consider  contributions to the $\Theta^+$ selfenergy from diagrams in which
the $\Theta^+$ couples to a nucleon and two mesons, like the one in Fig.
\ref{Threebody}.
There is no direct information on these couplings since the $\Theta^+$ mass is
below the two-meson decay threshold.

To proceed, we do several assumptions.
 First, the $\Theta^+$ is assumed to have $J^P=1/2^+$ associated to an
$SU(3)$ antidecuplet\cite{Diakonov:1997mm}. Also the
$N^*(1710)$ is supposed to couple strongly to the same antidecuplet.

From the PDG data on $N^*(1710)$ decays we  determine the couplings to the
two-meson channels, and using $SU(3)$ symmetry obtain the corresponding 
couplings for the $\Theta^+$ pentaquark.

 In order to account for the 
$N^*(1710)$ decay into $N (\pi\pi, p-\textrm{wave},I=1)$ and $N (\pi\pi,
s-\textrm{wave},I=0)$ we use the following lagrangians
\begin{equation}
\label{L1}
{\mathcal L}_1 = i g_{\bar{10}} \epsilon^{ilm} \bar{T}_{ijk} 
\gamma^{\mu} B^j_l (V_{\mu})^k_m \, ,
\end{equation}
with
\begin{equation}
\label{veccurr}
V_{\mu} = \frac{1}{4 f^2} (\phi \partial_{\mu} \phi - \partial_{\mu} 
\phi \phi)
\, ,
\end{equation}
where $f=93$ MeV is the pion decay constant and $T_{ijl}$, $B^j_l$, $\phi ^k_m$
$SU(3)$ tensors which account for the antidecuplet states, the octet of
$\frac{1}{2}^+$ baryons and the octet of $0^-$ mesons, respectively.
 The second term is given by
\begin{equation}
\label{L2}
{\mathcal L}_2 = \frac{1}{2 f} \tilde{g}_{\bar{10}}  \epsilon^{ilm} 
\bar{T}_{ijk}
(\phi \cdot \phi)^j_l B^k_m \, ,
\end{equation}
which couples two mesons in $L=0$ to the antidecuplet and the baryon and they
are in $I=0$ for the case of two pions.  From the Lagrangian terms of Eqs.
(\ref{L1}, \ref{L2}) we obtain the transition amplitudes 
$N^*\to \pi\pi N$. Taking the central values from the PDG\cite{PDG} for the
$N^* (1710) \to N(\pi\pi, p-\textrm{wave},I=1)$  and for the 
$N^* (1710) \to N(\pi\pi,s-\textrm{wave},I=0)$, the resulting coupling 
constants are $g_{\bar{10}}=0.72$
and  $\tilde{g}_{\bar{10}}=1.9$.

The implementation of the medium effects is done by including the medium
selfenergy of the kaon and modifying the nucleon propagator, as
before. On the other hand, for the pion we modify
the propagator using the $p-$wave selfenergy from $ph$ and $\Delta h$ 
excitations\cite{Oset:1981ih,Chiang:1997di,Cabrera:2004kt}.
Once the $\Theta^+$ selfenergy at a density $\rho$ is evaluated, the optical 
potential felt by the $\Theta^+$ in the medium is obtained by subtracting the
free $\Theta^+$ selfenergy.

We should also note that while the $\Theta^+\rightarrow K\pi N$ decay is
forbidden, in the medium the $\pi$ can lead to a $ph$ excitation and this opens
a new decay channel $\Theta^+ N\rightarrow NNK$, which is open down to 1432
MeV, quite below the free $\Theta^+$ mass. We have shown\cite{Cabrera:2004yg}
that the width from
this channel is also  very small, but should the $\Theta^+$ free width be of
the order of 1 MeV as suggested in  Refs. 
\refcite{Gibbs:2004ji,Sibirtsev:2004bg}, the new decay mode would make the
width in the medium larger than the free width.

\begin{figure}
 \begin{center} 
\includegraphics[width=0.85\textwidth]{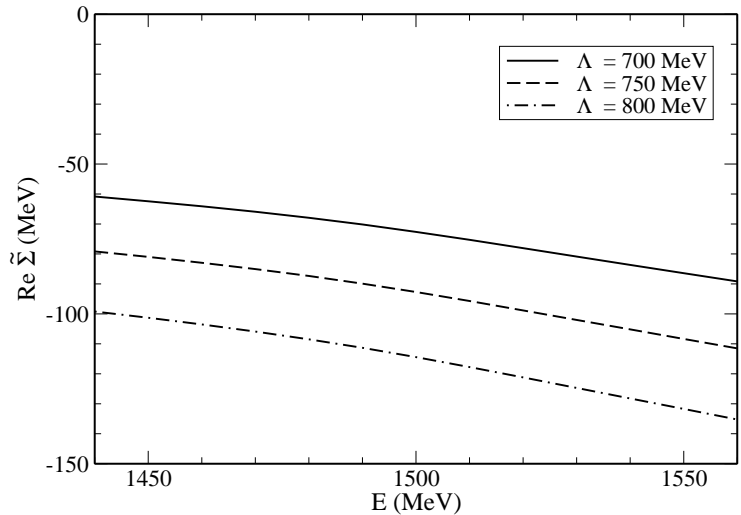}
\end{center} 
\caption{Real part of the two-meson contribution to the $\Theta^+$ selfenergy
at $\rho=\rho_0$.} 
\label{Re2meson}
\end{figure}

\begin{figure}
\begin{center}
\begin{center} 
    \includegraphics[width=0.85\textwidth]{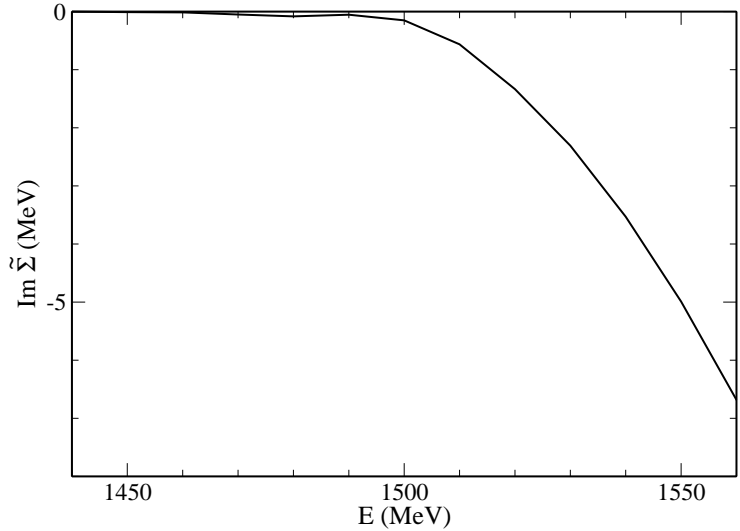}
\end{center} 
\caption{Imaginary part of the two-meson contribution to the $\Theta^+$ 
selfenergy at $\rho=\rho_0$.}
\label{Im2meson}
\end{center}
\end{figure}

We present the results in Figs. \ref{Re2meson} and \ref{Im2meson}. We can see
there that the potential for $\rho=\rho_0$ is sizable and attractive and goes
from $-70$ to $-120$ MeV depending on the cut-off used in the selfenergy
evaluation. 

Even with the quoted large uncertainties we conclude that there could be a 
sizable attraction of the order of magnitude of 50-100 MeV at normal nuclear 
density,
which is more than enough to bind the $\Theta^+$ in any nucleus. In Fig.
\ref{Im2meson} we show the imaginary part of the $\Theta^+$ selfenergy related
to the two-meson decay mechanism. We find that $\Gamma$ would be
smaller than 5 MeV for bound states with a binding of $\sim$20 MeV and
negligible for binding energies of $\sim$40 MeV or higher. This, together with
the small widths associated to the $KN$ decay channel, would lead to $\Theta^+$
widths below 8 MeV, assuming a free width of 15 MeV, and much lower if the
$\Theta^+$ free width is of the order of 1 MeV. In any case, for most
nuclei, this width would be smaller than the separation of the deep 
levels\cite{Nagahiro:2004wu}.

\section{Conclusions} 
The selfenergy of the $\Theta^+$ in the nuclear medium
associated to the $KN$ decay channels is quite weak, even assuming a  large
free width of around 15 MeV for the $\Theta^+$. However, there is a large
attractive $\Theta^+$ potential in the nucleus associated to the two meson
cloud of the antidecuplet. New decay channels open for the $\Theta^+$ in the
medium,  $\Theta^+ N\rightarrow NNK$, but the width from this new channel,
together with the one from $KN$ decay, is still small compared to the
separation of the bound levels of the $\Theta^+$ in light and intermediate
nuclei. 

This conclusions depend on several assumptions, namely:
 the $\Theta^+$ is assumed to be $1/2^+$ associated to an $SU(3)$ antidecuplet,
 the $N^*(1710)$ is supposed to couple largely to the same antidecuplet,
 two specific Lagrangians have been chosen,
 the average value of the $N^*(1710)$ width and the partial decay ratios, which
experimentally have large uncertainties, have been taken to fix the
couplings. 
It is clear that with all these assumptions one must accept a large uncertainty
in the results. So we can not be precise on the binding energies of the
$\Theta^+$. However, the order of magnitude obtained for the potential is such
that even with a wide margin of uncertainty, the conclusion that there would be
bound states is quite safe. In fact, with potentials with a strength of 20 MeV
or less one would already get bound states. Furthermore, since the strength of
the real part and the imaginary part from the $NKph$ decay are driven by the
same coupling, a reduction on the strength of the potential would also lead to
reduced widths such that the principle that the widths are reasonably smaller
than the separation between levels still holds.

\section*{Acknowledgments}
This work is partly supported by DGICYT contract number BFM2003-00856,
the E.U. EURIDICE network contract no. HPRN-CT-2002-00311 and by the
Research Cooperation Program of the Japan Society for the Promotion of Science
(JSPS) and the spanish Consejo Superior de Investigaciones Cientificas (CSIC).
D.~C. acknowledges financial support from MCYT and Q.~B.~Li acknowledges 
support from the Ministerio de Educaci\'on y Ciencia in the program of Doctores
y Tecn\'ologos Extranjeros.


\begin{thebibliography}{0}

\bibitem{Nakano:2003qx}
T.~Nakano {\it et al.}  [LEPS Collaboration],
{\it Phys.\ Rev.\ Lett.\ }  {\bf 91} (2003) 012002.

\bibitem{Miller:2004rj}
G.~A.~Miller,
{\it Phys.\ Rev.\ C} {\bf 70} (2004) 022202.


\bibitem{Kim:2004fk}
H.~C.~Kim, C.~H.~Lee and H.~J.~Lee,
arXiv:hep-ph/0402141.

\bibitem{Cabrera:2004yg}
D.~Cabrera, Q.~B.~Li, V.~K.~Magas, E.~Oset and M.~J.~Vicente Vacas,
arXiv:nucl-th/0407007.

\bibitem{madrid} 
T. Hyodo, F. Llanes-Estrada, E. Oset, J.R. Pelaez, A. Hosaka and
M. J. Vicente Vacas, in preparation.


\bibitem{Kaiser:1996js}
N.~Kaiser, T.~Waas and W.~Weise,
Nucl.\ Phys.\ A {\bf 612} (1997) 297.



\bibitem{Oset:2000eg}
E.~Oset and A.~Ramos,
Nucl.\ Phys.\ A {\bf 679} (2001) 616.



\bibitem{Cabrera:2002hc}
D.~Cabrera and M.~J.~Vicente Vacas,
Phys.\ Rev.\ C {\bf 67} (2003) 045203.

\bibitem{Diakonov:1997mm}
D.~Diakonov, V.~Petrov and M.~V.~Polyakov,
Z.\ Phys.\ A {\bf 359} (1997) 305.

\bibitem{PDG}
 S.~Eidelman et al., Phys.\ Lett.\ B {\bf 592} (2004) 1.

\bibitem{Oset:1981ih}
E.~Oset, H.~Toki and W.~Weise,
Phys.\ Rept.\  {\bf 83} (1982) 281.


\bibitem{Chiang:1997di}
H.~C.~Chiang, E.~Oset and M.~J.~Vicente Vacas,
Nucl.\ Phys.\ A {\bf 644} (1998) 77.


\bibitem{Cabrera:2004kt}
D.~Cabrera and M.~J.~Vicente Vacas,
Phys.\ Rev.\ C {\bf 69} (2004) 065204.


\bibitem{Gibbs:2004ji}
W.~R.~Gibbs,
arXiv:nucl-th/0405024.

\bibitem{Sibirtsev:2004bg}
A.~Sibirtsev, J.~Haidenbauer, S.~Krewald and U.~G.~Meissner,
arXiv:hep-ph/0405099.

\bibitem{Nagahiro:2004wu}
H.~Nagahiro, S.~Hirenzaki, E.~Oset and M.~J.~Vicente Vacas,
arXiv:nucl-th/0408002.

\end{thebibliography}
\end{document}